\batchmode
\documentclass[english,twocolumn,prl,showpacs]{revtex4}
\usepackage[latin1]{inputenc}
\usepackage{amsmath}
\usepackage{graphicx}

\usepackage{epsf}

\newcommand{\beq}{\begin{equation}}
\newcommand{\eeq}{\end{equation}}
\newcommand{\beqn}{\begin{eqnarray}}
\newcommand{\eeqn}{\end{eqnarray}}

\makeatother

\usepackage{babel}
\makeatother
\begin{document}

\title{Effective coupling between two Brownian particles}

\author{O. S. Duarte and A. O. Caldeira}

\affiliation{Departamento de F\'{\i}sica da Mat\'{e}ria Condensada, Instituto
de F\'{\i}sica Gleb Wataghin, Universidade Estadual de Campinas,
CEP 13083-970, Campinas-SP, Brazil}

\begin{abstract}
We use the system-plus-reservoir approach to study the dynamics of
a system composed of two independent Brownian particles. We
present an extension of the well-known model of a bath of
oscillators which is capable of inducing an effective coupling
between the two particles depending on the choice made for the
spectral function of the bath oscillators. The coupling is
non-linear in the variables of interest and an exponential
dependence on these variables is imposed in order to guarantee the
translational invariance of the model if the two particles are not
subject to any external potential. The effective equations of
motion for the particles are obtained by the Laplace transform
method and besides recovering all the local dynamical properties
for each particle we end up with an effective interaction
potential between them. We explicitly analyze one of its possible
forms.
\end{abstract}


\maketitle
Usually the Brownian motion of a given dynamical variable is modelled
by considering the system it describes - the system of interest -
coupled to a thermal bath responsible for its energy loss. Assuming
that any degree of freedom of this environment is only weakly perturbed
by the system of interest, we can describe it as a set of independent
harmonic oscillators with a coupling which is bilinear in the reservoir
and system variables and endowed with a particular spectral function
\cite{annals}. This model has been successfully used to describe
general properties, classical or quantum mechanical, of dissipative
systems with only one degree of freedom subject to arbitrary potentials
\cite{annals,physica,Leggett,Fisher,Garg}. Indeed, it has been extensively
shown in the literature that, within the range of interest, other
approaches to dealing with dissipative systems described by a single
dynamical variable always furnish us with the the same results as
those obtained by the bath of oscillators with a properly chosen spectral
function. It is the case , for example, of the application of the
collective coordinate method to describing the damped motion of quantum
solitons \cite{solitons} or microscopic attempts to describe more
realistic systems such as the electron gas of a metallic environment
\cite{Hedegard,Guinea,Weiss}.

However, despite all its success there are certain dissipative systems
for which the usual model of the bath of oscillators can be shown
to be inappropriate to account for the physics we expect from them.
Here it should be stressed that by the \textit{usual model} we mean
independent oscillators coupled \textit{bilinearly} in coordinates
to the system of interest.

Suppose one immerses two independent particles in the same medium
where each of them would separately behave as a Brownian particle.
Since for each individual particle we could mimic the effect of the
medium by the bath of oscillators it would be very natural to try
to generalize the model to cope with the presence of those two particles.
This generalization is quite straightforward and the only point where
one should be a bit cautious is when introducing the well-known counter-term
\cite{annals,physica} in the generalized model. In order to do that
in an unambiguous way all one has to do is employ the equivalent model
for the system-bath Lagrangian where the interaction is described
by a velocity-coordinate coupling and perform the usual canonical
transformations to achieve the desired coordinate-coordinate coupling
with the appropriate counter-term \cite{annals,physica}(see below).
It is then a simple matter to obtain the equations of motion for each
particle under the influence of the environment using, for example,
the Laplace transform method. These are coupled Langevin equations
that when written in terms of the center of mass and relative coordinates,
$q$ and $u$, of the two particles present a somewhat bizarre result;
namely, the latter obeys a free particle equation of motion. Therefore,
if we give an initial velocity to one of those particles the other
must instantaneously reply in such a way that the relative coordinate
evolves linearly in time ignoring completely the presence of the environment.
This is a very unexpected result to say the least.

It is our intention in this letter to propose an extension of the
usual model of a bath of oscillators in order to fix this deficiency.
As we will see, on top of succeeding in so doing, we will also be
able to describe an effective coupling between the two particles mediated
by the presence of the bath and the resulting interaction potential
depends on the specific form of the spectral function of the environment
oscillators. This effect reminds us of the formation of Cooper pairs
or bipolarons in material systems due to the electron-phonon interaction.

For the sake of completeness we shall start our procedure by briefly
reviewing some general aspects concerning the usual model for the
bath of oscillators that will be useful later on.

The Lagrangian for the complete system is given by \begin{equation}
L_{S}+L_{R}+L_{I}.\label{Ltotal}\end{equation}
 $L_{S}$ is the Lagrangian of the system of interest, $L_{R}$ is
the Lagrangian of the reservoir, which in the case of the bath of
oscillators is \begin{equation}
L_{R}=\sum\limits _{k=1}^{N}\left(\frac{m_{k}\dot{R}_{k}^{2}}{2}-\frac{m_{k}\omega_{k}^{2}R_{k}^{2}}{2}\right),\label{LR}\end{equation}
 and $L_{I}$ is the interaction Lagrangian which can be written in
two equivalent forms \cite{annals,physica}. For example, in an \textit{electromagnetic-like}
fashion we will have a coordinate-velocity coupling $L_{I}=-\sum\limits _{k}\tilde{C}_{k}\dot{R}_{k}x$.
Switching to the Hamiltonian formulation and performing the following
canonical transformation for the reservoir dynamical variables, $P_{k}\rightarrow m_{k}\omega_{k}R_{k}$
and $R_{k}\rightarrow\frac{P_{k}}{m_{k}\omega_{k}}$, we can show
that the coupling term transforms into \begin{equation}
L_{I}=-\sum_{k=1}^{N}\left(R_{k}C_{k}x+\frac{C_{k}^{2}}{2m_{k}\omega_{k}^{2}}x^{2}\right)\label{Lifin}\end{equation}
 where the coefficients have been redefined as $C_{k}\equiv\tilde{C}_{k}\omega_{k}$.
Therefore, the canonical transformation changes the coordinate-velocity
coupling into a coordinate-coordinate coupling plus a new quadratic
term which is necessary to preserve the translational invariance of
(\ref{Ltotal}) when the system of interest is not acted by any external
force. We refer the reader to \cite{Ambegaokar} for a discussion
on the invariant form of the latter.

In our generalization of the previous model, the system of interest
with a single degree of freedom will be represented by the free particle
Lagrangian \begin{eqnarray}
L_{S} & = & \frac{1}{2}M\dot{x}^{2}.\label{Ls}\end{eqnarray}

The heat bath will be described as a symmetrized collection of independent
harmonic modes, \begin{equation}
L_{R}=\frac{1}{2}\sum\limits _{k}m_{k}(\dot{R}_{k}\dot{R}_{-k}-\omega_{k}^{2}R_{k}R_{-k}).\label{LRS}\end{equation}

For the coupling term we initially assume the interaction Lagrangian
\begin{equation}
L_{I}=-\sum\limits _{k}\tilde{C}_{k}(x)\dot{R}_{k}.\label{LI-cv}\end{equation}

We have just shown above that this kind of coupling is equivalent
to the coordinate-coordinate coupling once an adequate potential renormalization
is introduced. Then, following the same procedure that led us to (\ref{Lifin})
it is easy to see the equivalence between (\ref{LI-cv}) and \begin{multline}
L_{I}=-\frac{1}{2}\sum_{k}\left(C_{-k}(x)R_{k}+C_{k}(x)R_{-k}\right)\\
-\sum_{k}\frac{C_{k}(x)C_{-k}(x)}{2m_{k}\omega_{k}^{2}}.\label{LI-cc}\end{multline}
 where $C_{k}(x)\equiv\tilde{C}_{k}(x)\omega_{k}$. In order to represent
the effect of a local interaction of the particle with a spatially
homogeneous environment we choose \begin{eqnarray}
C_{k}(x) & = & \kappa_{k}e^{ikx}.\label{acoplamento}\end{eqnarray}

With this choice it is easy to show that the entire system is translationally
invariant and, moreover, the coupling to the environment is homogeneous.
If the particle is displaced by a distance $d$, the coupling term
transforms into $C_{-k}(x+d)R_{k}=C_{-k}(x)e^{-ikd}R_{k}$, which
suggests the definition of a new set of canonical variables as $\tilde{R}_{k}=e^{-ikd}R_{k}$
that renders the total Lagrangian invariant. It is important to notice
that with a coupling like (\ref{acoplamento}), the potential renormalization
in (\ref{LI-cc}) is a constant and therefore does not contribute
to the particle dynamics. A coupling like (\ref{acoplamento}) appears,
for example, in the interaction of a particle with the density operator
of a fermionic bath \cite{Hedegard}.

Now, we write the Euler-Lagrange equations for all the variables involved
in the problem, take their Laplace transforms and eliminate the bath
variables, in order to find the following equation of motion for the
system of interest \begin{equation}
M\ddot{x}+\int_{0}^{t}K(x(t)-x(t'),t-t')\dot{x}(t')dt'=F(t).\label{eq
mov}\end{equation}
 Here the nonlinear dissipation kernel, given by \begin{equation}
K(r,\tau)=\sum_{k}\frac{k^{2}\kappa_{k}\kappa_{-k}}{m_{k}\omega_{k}^{2}}\cos kr\cos\omega_{k}\tau,\label{Kernell}\end{equation}
 shows that the interaction with the thermal bath induces a systematic
influence on the system which is non-local and non-instantaneous.
The function $F(t)$ can be interpreted as a fluctuating force\begin{align}
 & F(t)=-\frac{\partial}{\partial x}\sum\limits _{k}\biggl\{\left(C_{-k}(x)\tilde{R}_{k}(0)+C_{k}(x)\tilde{R}_{-k}(0)\right)\frac{\cos\omega_{k}t}{2}\nonumber \\
 & +\left(C_{-k}(x)\dot{R}_{k}(0)+C_{k}(x)\dot{R}_{-k}(0)\right)\frac{\sin\omega_{k}t}{2\omega_{k}}\biggr\},\label{f(t)}\end{align}
where $\tilde{R}_{k}=R_{k}+\frac{C_{k}(x_{0})}{m_{k}\omega_{k}^{2}}$
. $F(t)$ depends explicitly on the initial conditions of the bath
variables and its statistical properties are obtained from the initial
state of the total system. Now we need to choose a suitable distribution
of oscillators, in the continuum limit, which leads us to the Brownian
motion. The kernel in (\ref{Kernell}) can be written as\begin{multline}
K=\sum_{k}\int_{0}^{\infty}d\omega2k^{2}\kappa_{k}\kappa_{-k}\frac{\textrm{Im}\chi_{k}^{(0)}(\omega)}{\pi\omega}\\
\cos k(x(t)-x(t'))\cos\omega(t-t'),\label{TD2}\end{multline}
where $\textrm{Im}\chi_{k}^{(0)}(\omega)=\frac{\pi}{2m_{k}\omega_{k}}\delta(\omega-\omega_{k})$
is the imaginary part of the dynamical response $\chi_{k}^{(0)}(\omega)$
of a non-interacting oscillator with wave number $k$. If we now assume
that the bath oscillators are in fact only approximately non-interacting
and replace \textit{only} their response functions by those of damped
oscillators, $\chi_{k}(\omega)$, one has\begin{equation}
\textrm{Im}\chi_{k}(\omega)=\frac{\gamma_{k}\omega}{m_{k}\left[\left(\omega^{2}-\omega_{k}^{2}\right)^{2}+\omega^{2}\gamma_{k}^{2}\right]},\label{chiin}\end{equation}
 where $\gamma_{k}$ is the relaxation frequency of that oscillator.

Our main interest is to study the system for times longer than the
typical time scale of the reservoir. In other words, we wish to study
the low-frequency limit of eq. (\ref{chiin}) in which $\textrm{Im}\chi_{k}(\omega)\propto\omega$
and, therefore, we can assume \begin{equation}
\textrm{Im}\chi_{k}(\omega)\approx f(k)\omega\theta(\Omega-\omega),\label{chiIntempos}\end{equation}
 where $f(k)\approx\frac{\gamma_{k}}{m_{k}\omega_{k}^{4}}$. Here
we have, as usual, introduced a high frequency cutoff $\Omega$ as
the characteristic frequency of the bath. A functional dependence
like (\ref{chiIntempos}) for the dynamical response of the bath
has been employed in the references \cite{Hedegard,Guinea} for
fermionic environments. The particular choice of the dynamical
susceptibility of the bath allow us to separate its time and
length scales and to obtain a Markov dynamics when we replace
(\ref{chiIntempos}) in (\ref{TD2}) and integrate with respect to
$\omega$ taking the limit $\Omega\rightarrow\infty$. With these
considerations the equation of motion (\ref{eq mov}) reads
\begin{equation}
M\ddot{x}(t)+\eta\dot{x}(t)=F(t),\label{eq lang NL}\end{equation}
 where we have defined $\eta=\sum_{k}k^{2}\kappa_{k}\kappa_{-k}f(k)$.
Notice that with this modification we obtain a relation between the
damping constant and some microscopic parameters of the oscillator
bath. With the current prescription and supposing that the bath is
initially in thermal equilibrium it is easy to show that, for \textit{high
temperatures}, the fluctuating force $F(t)$ satisfies the relations
$\left\langle F(t)\right\rangle =0$ and $\left\langle F(t)F(t')\right\rangle =2\eta k_{B}T\delta(t-t')$,
which are characteristics of white noise. Notice that this is valid
only if we assume a classical distribution of oscillators as the initial
state of the bath.

In conclusion, the system-reservoir model with non-linear coupling
presented here allows us to reproduce the result one would have
obtained by coupling the particle of interest bilinearly to a bath
of non-interacting harmonic oscillators with the spectral function
$J(\omega)=\eta\omega$ \cite{annals,physica}.

Now we are going to study the dynamics of a system with two degrees
of freedom immersed in a dissipative environment. In this case the
Lagrangian of the system of interest is \begin{equation}
L_{S}=\frac{1}{2}M\dot{x}_{1}^{2}+\frac{1}{2}M\dot{x}_{2}^{2},\label{LS-2P}\end{equation}
 and the coupling term \begin{multline}
L_{I}=-\frac{1}{2}\sum_{k}\left[\left(C_{-k}(x_{1})+C_{-k}(x_{2})\right)R_{k}\right.\\
\left.+\left(C_{k}(x_{1})+C_{k}(x_{2})\right)R_{-k}\right].\label{LI-2P}\end{multline}

Notice that we have not included any counter-term in (\ref{LI-2P})
since our system is translationally invariant. The equations of
motion for this Lagrangian are then \begin{align}
M\ddot{x}_{i} & +\int_{0}^{t}K(x_{i}(t)-x_{i}(t'),t-t')\dot{x}_{i}(t')dt'\nonumber \\
 & +\int_{0}^{t}K(x_{i}(t)-x_{j}(t'),t-t')\dot{x}_{j}(t')dt'\nonumber \\
 & +\frac{\partial}{\partial x_{i}}V(x_{i}(t)-x_{j}(t))=F_{i}(t),\label{eq mov 2P}\end{align}
 where $i\neq j=1,2$, the fluctuating force $F_{i}(t)$ has the form
given by (\ref{f(t)}) replacing $x(t)$ by $x_{i}(t)$ and \begin{equation}
V(r(t))=-\sum_{k}\frac{\kappa_{k}\kappa_{-k}}{m_{k}\omega_{k}^{2}}\cos kr(t).\label{Pefe}\end{equation}

In the long-time limit, the second term in (\ref{eq mov 2P}) can
be written as $\eta\dot{x}_{i}$ providing us with the usual dissipative
term. The third term represents a cross-dissipative term that depends
on the velocity of the second particle and the relative distance between
them. The last term is clearly an effective interaction induced by
the coupling with the thermal reservoir. To see the explicit form
of this interaction and the non-local influence in the cross-dissipative
term, we need to evaluate the sum in (\ref{TD2}) with $\textrm{Im}\chi_{k}^{(0)}(\omega)$
replaced by (\ref{chiIntempos}). For this we had better define the
function $g(k)$ as \begin{equation}
\eta g(k)=\frac{L}{2\pi}\kappa_{k}\kappa_{-k}f(k),\label{g(k)}\end{equation}
 where we have assumed a one-dimensional environment of length $L$.

The exact form of this function can be obtained from a microscopic
description of the bath and the details of the interaction between
the system of interest and the bath degrees of freedom. As we are
not interested in such a detailed description of the many-body reservoir,
we are going to choose a form for $g(k)$ that satisfies the condition
$\int_{0}^{\infty}g(k)k^{2}dk=1$ derived from the definition of the
damping constant $\eta$. Notice that we have replaced $\sum_{k}\rightarrow\frac{L}{2\pi}\int dk$.
A reasonable choice for this function is \begin{equation}
g(k)=Ae^{-k/k_{0}},\label{g(k)2}\end{equation}
 where $A=1/(2k_{0}^{3})$ is the normalization constant and $k_{0}$
determines the characteristic length scale of the environment. For
example, in the case of fermionic environments $k_{0}$ is of the
order of $k_{F}$ \cite{Hedegard}. One should bear in mind that the
choice of $g(k)$ is guided either by the knowledge of the microscopic
details of the environment or by some phenomenological input about
the effective interaction between the two particles.

With the form chosen for $g(k)$ the kernel of the third term in (\ref{eq mov 2P})
is, again only for \textit{high temperatures}, \begin{equation}
K(x_{1}(t)-x_{2}(t'),t-t')=2\delta(t-t')\eta(u(t)),\label{kernell
mark}\end{equation}
 where $u(t)=x_{1}(t)-x_{2}(t)$ and \begin{equation}
\eta(u(t))=\eta\left(\frac{1}{\left(k_{0}^{2}u^{2}+1\right)^{2}}-\frac{4u^{2}k_{0}^{2}}{\left(k_{0}^{2}u^{2}+1\right)^{3}}\right),\label{eta(u)}\end{equation}
 and the effective potential reads \begin{equation}
V(u(t))=-\frac{2\Omega\eta}{\pi k_{0}^{2}\left(k_{0}^{2}u^{2}(t)+1\right)}.\label{V(u)}\end{equation}

The strength of the effective potential depends on the characteristic
length and time scales. Therefore the contribution of this coupling
to the dynamics of the Brownian particles is only important for times
longer than the typical time scale of the reservoir and distances
of the order of (or less than) the characteristic length $k_{0}^{-1}$.
The fluctuating forces still satisfy the white noise properties but
now they present an additional property associated with the distance
between the particles. The forces $F_{1}(t)$ and $F_{2}(t)$ are
also spatially correlated, that is \begin{align}
\left\langle F_{1}(t)F_{2}(t')\right\rangle = & 2\eta(u(t))k_{B}T\delta(t-t').\label{correl f1 f2}\end{align}

\begin{figure}
\begin{centering}\includegraphics{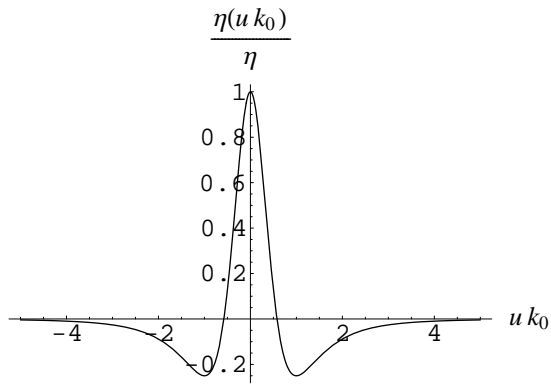}\par\end{centering}

\caption{\label{cap:Spatial-Correlation-of}Spatial Correlation of $F_{1}$
and $F_{2}$}
\end{figure}

In figure (\ref{cap:Spatial-Correlation-of}) we see the noise correlation
strength as a function of the distance between the particles. For
short distances the noise correlation has the standard Brownian behavior.
However, for longer distances the correlation function becomes negative
and this \textit{anti-correlation} induces an anomalous diffusive
process in the system which will ultimately tend to normal diffusion
once the particles are infinitely far apart.

In terms of the relative and center-of-mass coordinates the equations
of motions read \begin{equation}
M\ddot{u}(t)+\eta\dot{u}(t)-\eta(u(t))\dot{u}+V'(u(t))=F_{u}(t)\label{u(t)}\end{equation}
 and \begin{equation}
M\ddot{q}(t)+\eta\dot{q}(t)+\eta(u(t))\dot{q}=F_{q}(t),\label{q(t)}\end{equation}
 where $F_{u}(t)=F_{1}(t)-F_{2}(t)$ and $F_{q}(t)=(F_{1}(t)+F_{2}(t))/2$.
>From the form of $V(u(t))$ and $\eta(u(t))$ and the statistical
properties of the fluctuating forces, $\langle F_{u}(t)\rangle=\langle F_{q}(t)\rangle=0$,
$\langle F_{u}(t)F_{u}(t')\rangle=4kT(\eta-\eta(u))\delta(t-t')$
and $\langle F_{q}(t)F_{q}(t')\rangle=kT(\eta+\eta(u))\delta(t-t')$,
it is evident that at large distances the equations of motion for
the relative and center of mass coordinates represent the motion of
Brownian particles with reduced mass $M/2$ and total mass $2M$,
respectively. In general the dissipation depends on the relative distance
between the particles and for distances such that $uk_{0}\ll1$, we
have up to first order in $u(t)$, $V'(u)\propto-u(t)$. In this approximation
both dissipation and fluctuations are negligible and then we have
an undamped oscillatory motion for $u(t)$.

In conclusion, we presented a system-plus-reservoir model with a
coupling which is non-linear in the system coordinates and, in the
adequate limit, allows us to reproduce the phenomenological
results known for the dynamics of a dissipative system with only
one degree of freedom. In our model the dissipation coefficient is
directly expressed in terms of a few parameters of the thermal
bath that can be measured experimentally. Moreover, our model is
capable of inducing an effective coupling between the two Brownian
particles which arises from the non-local effects generated by the
bath and depends on the choice of the dynamical susceptibility of
the reservoir. In the single particle case, the non-local effects
will be relevant only in a non-markovian approximation.

We should finally stress that we have only reported our findings
for the classical or high temperature regime. The quantum
mechanical analysis of this system requires the evaluation of the
reduced density operator for the two particles of interest down to
very low temperatures and will be presented elsewhere
\cite{Oscar}.

We kindly acknowledge Conselho Nacional de Desenvolvimento Cient\'{\i}fico
e Tecnol\'{o}gico (CNPq) for the financial support.

\end{document}